\documentclass{sig-alternate}
\usepackage{mathptmx}
\usepackage{comment}

\usepackage{graphicx}
\usepackage{epsfig}
\usepackage{cite}
\usepackage{fancyhdr}
\usepackage[normalem]{ulem}
\usepackage[hyphens]{url}
\usepackage{hyperref}
\usepackage[utf8]{inputenc}
\usepackage{tikz}

\newcommand\encircle[1]{%
  \tikz[baseline=(X.base)]
  \node (X) [draw,
    shape=circle, inner sep=0]
  {\strut #1};}

\title{Hardware Translation Coherence for Virtualized Systems \vspace{-8mm}}

\author{Zi Yan, Guilherme Cox, J{\'a}n Vesel{\'y}, Abhishek Bhattacharjee\\
{ Department of Computer Science}\\
{ \{zi.yan, guilherme.cox, jan.vesely, abhib\}@cs.rutgers.edu}}

\begin{document}
\maketitle
\pagestyle{plain}

\begin{abstract}

To improve system performance, modern operating systems (OSes) often
undertake activities that require modification of virtual-to-physical
page translation mappings. For example, the OS may migrate data
between physical frames to defragment memory and enable
superpages. The OS may migrate pages of data between heterogeneous
memory devices. We refer to all such activities as page
remappings. Unfortunately, page remappings are expensive. We show that
translation coherence is a major culprit and that systems employing
virtualization are especially badly affected by their overheads. In
response, we propose \textsf{\textbf{hardware translation invalidation
    and coherence}} or \textsf{\textbf{HATRIC}}, a readily
implementable hardware mechanism to piggyback translation coherence
atop existing cache coherence protocols. We perform detailed studies
using KVM-based virtualization, showing that {\sf HATRIC} achieves up
to 30\% performance and 10\% energy benefits, for per-CPU area
overheads of 2\%. We also quantify {\sf HATRIC's} benefits on systems
running Xen and find up to 33\% performance improvements.

\end{abstract}

\section{Introduction}\label{introduction}
As the computing industry designs systems for big-memo\-ry workloads,
systems architects have begun embracing heterogeneous memory
architectures. For example, Intel is integrating high-bandwidth
on-package memory in its Knight's Landing chip, and 3D Xpoint memory
in several products \cite{intel:3dxpoint}. AMD and Hynix are releasing
High-Bandwidth Memory or HBM \cite{kim:hbm, black:die}. Similarly,
Micron's Hybrid Memory Cube \cite{micron:hmc, pawlowski:hybrid} and
byte-addressable persistent memories \cite{xie:emerging,
  xie:emerging2, dong:circuit, ramos:page} are quickly gaining
traction. Vendors are combining these high-performance memories with
traditional high-capacity and low-cost DRAM, prompting research on
heterogeneous memory architectures \cite{ramos:page, oskin:sw,
  kim:hbm, phadke:mlp, meswani:hetero, ausavarungnirun:staged,
  agarwal:page, vesely:observations}.

Fundamentally, heterogeneous memories are dependent on the concept of
page remapping to migrate data between diverse memory devices for good
performance. Page remapping is not a new concept -- OSes have long
used it to migrate physical pages to defragment memory and create
superpages \cite{arcangeli:ths, navarro:practical, talluri:surpassing,
  kwon:coordinated}, to migrate pages among NUMA sockets
\cite{gaud:numa, lepers:thread}, and to deduplicate memory by enabling
copy-on-write optimizations \cite{pham:glue, pham:tr,
  seshadri:page}. However, while page remappings were used sparingly
in those scenarios, they are likely to be used more frequently for
heterogeneous memories. This is because page remapping is essential to
adapt data placement to the memory access patterns of workloads, and
to harness the performance and energy potential of memories with
different latency, bandwidth, and capacity
characteristics. Consequently, developers at IBM and Redhat are
already deploying Linux patchsets to enable page remapping amongst
coherent heterogeneous memory devices \cite{ibm:lwn, redhat:lwn1,
  redhat:lwn2}.

Unfortunately, these efforts face an obstacle -- the high performance
and energy penalty of page remapping. There are two components to this
cost. The first is the overhead of copying data. The second is the
cost of translation coherence. When privileged software remaps a
physical page, it has to update the corresponding virtual-to-physical
page translation in the page table. Translation coherence is the means
by which caches dedicated to translations (e.g., Translation Lookaside
Buffers or TLBs \cite{pham:CoLT, pham:clustering, bhattacharjee:tlb,
  lustig:tlb}, etc.) are kept up to date with the latest page table
mappings.

Past work has shown that translation coherence overheads can easily
consume 10-30\% of system performance \cite{oskin:sw,
  romanescu:unified, villavieja:didi}. These overheads are even more
alarming on virtualized systems, which are used in the server and
cloud settings expected to be early adopters of heterogeneous
memories. We are the first to show that as much as 40\% of their
runtime can be wasted on translation coherence. The key culprit is
virtualization's use of multiple page tables. Architectures with
hardware assists for virtualization like Intel VT-x and AMD-V use a
guest page table to map guest virtual pages to guest physical pages,
and a nested page table to map guest physical pages to system physical
pages. Changes to the guest page table and in particular, the nested
page table, prompt expensive translation coherence activity.

The problem of coherence is not restricted to translation mappings. In
fact, the systems community has studied problems posed by cache
coherence for several decades \cite{sorin:primer} and has developed
efficient hardware cache coherence protocols \cite{martin:cc}. What
makes translation coherence challenging is that unlike cache
coherence, it relies on cumbersome software support. While this may
have sufficed in the past when page remappings were used relatively
infrequently, they are problematic for heterogeneous memories where
page remapping is more frequent. Consequently, we believe that there
is a need to architect better support for translation coherence. In
order to understand what this support should constitute, we list three
attributes desirable for translation coherence.

\vspace{2mm}\noindent \textcircled{\small 1} {\bf Precise
  invalidation:} Processors use several hardware translation
structures -- TLBs, MMU caches \cite{barr:translation,
  bhattacharjee:mmu}, and nested TLBs (nTLBs) \cite{bhargava:asplos}
-- to cache portions of the page table(s). Ideally, translation
coherence should invalidate the translation structure entries
corresponding to remapped pages, rather than flushing all the contents
of these structures.

\vspace{2mm}\noindent \textcircled{\small 2} {\bf Precise target
  identification:} The CPU running privileged code that remaps a page
is known as the {\it initiator}. An ideal translation coherence
protocol would allow the initiator to identify and alert only CPUs
whose TLBs, MMU caches, and nTLBs actually cache the remapped page's
translation. By restricting coherence messages to only these {\it
  targets}, other CPUs remain unperturbed by coherence activity.

\vspace{2mm}\noindent \textcircled{\small 3} {\bf Lightweight
  target-side handling:} Target CPUs should invalidate their
translation structures and relay acknowledgment responses to the
initiator quickly, without excessively interfering with workloads
executing on the target CPUs.

\vspace{2mm} Unfortunately, translation coherence meets none of these
goals today. Consider, for example, changes to the nested page
table. Further, consider \textcircled{\small 1}; when hypervisors
change a nested page table entry, they track guest physical and system
physical page numbers, but not the guest virtual page. Unfortunately,
as we describe in Sec. \ref{background}, translation structures on
architectures like x86-64 permit invalidation of individual entries
only if their guest virtual page is known. Consequently, hypervisors
completely flush all translation structures, even when only a single
page is remapped. This degrades performance since virtualized systems
need expensive two-dimensional page table walks to re-populate the
flushed structures \cite{pham:glue, bhargava:asplos, chang:improving,
  ahn:revisiting, gandhi:agile, cox:efficient,
  bhattacharjee:translation, pham:CoLT, pham:clustering}.

Current translation coherence protocols also fail to achieve
\textcircled{\small 2}. Hypervisors track the subset of CPUs that a
guest VM runs on but cannot (easily) identify the CPUs used by a
process within the VM. Therefore, when the hypervisor remaps a page,
it conservatively initiates coherence activities on all CPUs that may
potentially have executed {\it any} process in the guest VM. While
this does spare CPUs that never execute the VM, it needlessly flushes
translation structures on CPUs that execute the VM but not the
process.

Finally, \textcircled{\small 3} is also not met. Initiators currently
use expensive inter-processor interrupts (on x86) or {\sf tlbi}
instructions (on ARM, Power) to prompt VM exits on all target
CPUs. Translation structures are flushed on a VM re-entry. VM exits
are particularly detrimental to performance, interrupting the
execution of target-side applications \cite{bhargava:asplos,
  adams:comparison}.

We believe that the solution to these problems is to implement
translation coherence in hardware. This view is inspired by prior work
on {\sf UNITD} \cite{romanescu:unified}, which showcased the potential
of hardware translation coherence. Unfortunately, {\sf UNITD} is
energy inefficient and, like other recent proposals \cite{oskin:sw,
  villavieja:didi}, cannot support virtualized systems. In response,
we propose {\bf \textsf{\underline{ha}rdware \underline{tr}anslation
    \underline{i}nvalidation and \underline{c}oh\-e\-rence}} or {\bf
  \textsf{HATRIC}}, a hardware mechanism to tackle these problems and
meet \textcircled{\small 1}-\textcircled{\small 3}. {\sf HATRIC}
extends translation structure entries with coherence tags (or co-tags)
storing the system physical address where the translation entry
resides (not to be confused with the physical address stored in the
page table). This solves \textcircled{\small 1}, since translation
structures can now be identified by the hypervisor {\it without}
knowledge of the guest virtual address. {\sf HATRIC} exposes co-tags
to the underlying cache coherence protocol, achieving
{\textcircled{\small 2}} and {\textcircled{\small 3}}.

We evaluate {\sf HATRIC} for a forward-looking virtualized system with
a high-bandwidth die-stacked memory and a slower off-chip memory. {\sf
  HATRIC} drastically reduces translation coherence overheads,
improving performance by 30\%, saving as much as 10\% of energy, while
adding less than 2\% of CPU area. Overall, our contributions are:

\vspace{2mm}
\begin{itemize}
\item We perform a characterization study to quantify the overheads of
  translation coherence on hypervisor-man\-aged die-stacked
  memory. While we focus on KVM in this paper, we have also studied
  Xen and quantified its overheads.

\vspace{2mm}
\item We design {\sf HATRIC} to subsume translation coherence in
  hardware by piggybacking on, without fundamentally changing,
  existing cache coherence protocols. {\sf HATRIC} goes beyond {\sf
    UNITD} \cite{romanescu:unified} by \textcircled{\small a}
  accommodating translation coherence for both bare-metal {\it and}
  virtualized scenarios; \textcircled{\small b} extending coherence to
  not just TLBs, but also MMU caches and nTLBs; \textcircled{\small c}
  and achieving better energy efficiency.

\vspace{2mm}
\item We perform several studies that illustrate the benefits of {\sf
  HATRIC's} design decisions. Further, we discuss {\sf HATRIC's}
  advantages over purely software approaches to mitigate translation
  coherence issues.

\end{itemize}
 
\vspace{2mm} Overall, {\sf HATRIC} is efficient and versatile. While
we mostly focus on the particularly arduous challenges of translation
coherence due to nested page table changes, {\sf HATRIC} is applicable
to guest page tables and non-virtualized systems.

\section{Background}\label{background}
We begin by presenting an overview of the key hardware and software
structures involved in page remapping. Our discussion focuses on
x86-64 systems. Other architectures are broadly similar but differ in
some low-level details.

\subsection{HW and SW Support for Virtualization}\label{hw-sw-virtualization}

Virtualized systems accomplish virtual-to-physical address translation
in one of two ways. Traditionally, hypervisors have used shadow page
tables to map guest virtual pages (GVPs) to system physical pages
(SPPs), keeping them synchronized with guest OS page tables
\cite{ahn:revisiting}. However, the overheads of page table
synchronization can often be high \cite{gandhi:agile}. As a result,
most modern systems now use two dimensional page tables
instead. Figure \ref{pt-translation-structures} illustrates
two-dimensi\-o\-nal page table walks (see past work for more details
\cite{pham:glue, barr:translation, bhargava:asplos, ahn:revisiting,
  gandhi:efficient, barr:spectlb, cox:efficient}). Guest page tables
map GVPs to guest physical pages (GPPs). Nested page tables map GPPs
to SPPs. x86-64 systems use 4-level forward mapped radix trees for
both page tables \cite{pham:glue, bhargava:asplos, barr:spectlb,
  gandhi:efficient}. We refer to these as levels 4 (the root level) to
1 (the leaf level) as per recent work \cite{bhargava:asplos,
  bhattacharjee:mmu, barr:translation}. When a process running in a
guest VM makes a memory reference, its GVP must be translated to an
SPP. Consequently, the guest {\sf CR3} register is combined with the
requested GVP (not shown in the picture) to deduce the GPP of level 4
of the guest page table (shown as {\sf GPP Req.}). However, to look up
the guest page table ({\sf gL4-gL1}), the GPP must be converted into
the SPP where the page table actually resides. Therefore, we first use
the GPP to look up the nested page tables ({\sf nL4-nL1}), to find SPP
{\sf gL4}. Looking up {\sf gL4} then yields the GPP of the next guest
page table level ({\sf gL3}). The rest of the page table walk proceeds
similarly, requiring 24 memory references in total. This presents a
performance problem as the number of references is significantly more
than the 4 references needed for non-virtualized systems. Further, the
references are entirely sequential. CPUs use three types of
translation structures to accelerate this walk:

\begin{figure}[t]
\centering
{
\begin{minipage}[t]{0.48\textwidth}
\centering
\vspace{-6mm}
\epsfig{file=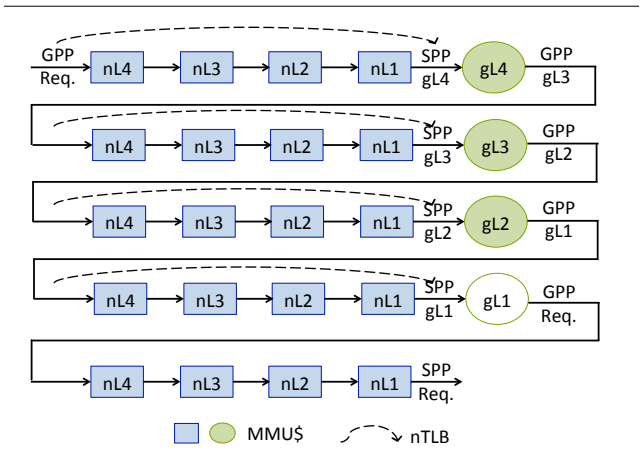, scale=0.34, angle=90}
\vspace{-14mm}
\caption{\small Two-dimensional page table walks for virtualized
  systems. Nested page tables are represented by boxes and guest page
  tables are represented by circles. Each page table's levels from 4
  to 1 are shown. We show items cached by MMU caches and nTLBs. TLBs
  (not shown) cache translations from the requested guest virtual page
  (GVP) to the requested system physical page (SPP).}
\label{pt-translation-structures}
\end{minipage}
}
\end{figure}

\vspace{2mm}\noindent{\bf \textcircled{a} Private per-CPU TLBs} cache
the requested GVP to SPP mappings, short-circuiting the entire
walk. TLB misses trigger hardware page table walkers to look up the
page table.

\vspace{2mm}\noindent{\bf \textcircled{b} Private per-CPU MMU caches}
store intermediate page table information to accelerate parts of the
page table walk \cite{barr:translation, bhattacharjee:mmu,
  bhargava:asplos}. There are two flavors of MMU cache. The first is a
{\it page walk cache} and is implemented in AMD chips
\cite{bhattacharjee:mmu, bhargava:asplos}. Figure
\ref{pt-translation-structures} shows the information cached in page
walk caches. Page walk caches are looked up with GPPs and provide SPPs
where page tables are stored. The second is called a {\it paging
  structure cache} and is implemented by Intel \cite{bhattacharjee:mmu,
  barr:translation}. Paging structure caches are looked up with GVPs
and provide the SPPs of page table locations. Paging structure caches
generally perform better, so we focus mostly on them
\cite{bhattacharjee:mmu, barr:translation}.

\vspace{2mm}\noindent{\bf \textcircled{c} Private per-CPU nTLBs}
short-circuit nested page table lookups by caching GPP to SPP
translations \cite{bhargava:asplos}. Figure
\ref{pt-translation-structures} shows the information cached by nTLBs.

\vspace{2mm} Concomitantly, CPUs cache page table information in
private L1 (L2, etc.) caches and the shared last-level cache
(LLC). The presence of separate private translation caches poses
coherence problems. While standard cache coherence protocols ensure
that page table entries in private L1 caches are coherent, there are
no such guarantees for TLBs, MMU caches, and nTLBs. Instead,
privileged software keeps translation structures coherent with data
caches and one another.

\subsection{Page Remapping in Virtualized Systems}\label{page-remapping-taxonomy}

We now detail the ways in which a virtualized system can trigger
coherence activity in translation structures. All page remappings can
be classified by the data they move, and the software agent initiating
the move.

\vspace{2mm}\noindent {\bf Remapped data:} Systems may remap a page
storing (i) the guest page table; (ii) the nested page table; or (iii)
non-page table data. Most remappings are from (iii) as they constitute
most memory pages. We have found that less than 1\% of page remappings
correspond to (i)-(ii). We therefore highlight {\sf HATRIC's}
operation using (iii); nevertheless, {\sf HATRIC} also implicitly
supports the first two cases.

\vspace{2mm}\noindent {\bf Remapping initiator:} Pages can be remapped
by (i) a guest OS; or (ii) the hypervisor. When a guest OS remaps a
page, the guest page table changes. Past work achieves low-overhe\-ad
guest page table coherence with relatively low-complexity software
extensions \cite{ouyang:shoot4u}. Unfortunately, there are no such
workarounds to mitigate the translation coherence overheads of
hypervisor-initiated nested page table remappings. For these reasons,
cross-VM memory deduplication \cite{pham:glue, guo:proactively} and
page migration between NUMA memories on multi-socket systems
\cite{rao:vnuma-mgr, rao:optimizing, banerjee:numa} are known to be
expensive. In the past, such overheads may have been mitigated by
using these optimizations sparingly. However, nested page table
remappings become frequent with heterogeneous memories, making
hypervisor-initiated translation coherence problematic.

\section{Shortcomings of Current Translation Coherence Mechanisms}\label{motivation}

Our goal is to ensure that translation coherence does not impede the
adoption of heterogeneous memories. We study forward-looking
die-stacked DRAM as an example of an important heterogeneous memory
system. Die-stacked memory uses DRAM stacks that are tightly
integrated with the processor die using high-bandwidth links like
through-silicon vias, or silicon interposers \cite{oskin:sw,
  kannan:enabling}.  Die-stacked memory is expected to be useful for
multi-tenant and rack-scale computing where memory bandwidth is often
a performance bottleneck, and will require a combination of
application, guest OS, and hypervisor management \cite{oskin:sw,
  falsafi:rack, vmware:best, loh:supporting}. We take the first steps
towards this, by showing the problems posed by translation coherence
on hypervisor management.

\subsection{Translation Coherence Overheads}\label{translation-coherence-overheads}

We quantify translation coherence overheads on a die-stacked system
that is virtualized with KVM. We modify KVM to page between the
die-stacked and off-chip DRAM. Since ours is the first work to
consider hypervisor management of die-stacked memory, we implement a
variety of paging policies. Rather than focusing on developing a
single ``best'' policy, our objective is to show that current
translation coherence overheads are so high that they curtail the
effectiveness of practically any paging policy.

Our paging mechanisms extend prior work that explores basic
software-guided die-stacked DRAM paging \cite{oskin:sw}. When off-chip
DRAM data is accessed, there is a page fault. KVM then migrates the
desired page into an available die-stacked DRAM physical page
frame. The GVP and GPP remain unchanged, but KVM changes the SPP and
hence, its nested page table entry. This triggers translation
coherence.

We run our modified KVM on the detailed cycle-accurate simulator
described in Sec. \ref{methodology}. Like prior work \cite{oskin:sw},
we model a system with 2GB of die-stacked DRAM with 4$\times$ the
memory bandwidth of a slower off-chip 8GB DRAM. This is a total of
10GB of addressable DRAM. Further, we model 16 CPUs based on Intel's
Haswell architecture.

\begin{figure}[t]
\centering
{
\begin{minipage}[t]{0.48\textwidth}
\centering
\vspace{-4mm}
\epsfig{file=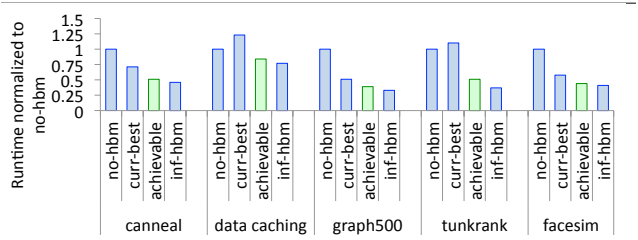, scale=0.34, angle=90}
\vspace{-42mm}
\caption{\small Performance of {\sf no-hbm} (no die-stacked DRAM),
  {\sf inf-hbm} (data always in die-stacked DRAM), {\sf curr-best}
  (best die-stacked DRAM paging policy with current software
  translation coherence overheads), and {\sf achievable} (best
  achievable paging policy, assuming no translation coherence
  overheads). All data is normalized to {\sf no-hbm} runtime.}
\vspace{-4mm}
\label{motivation-perf}
\end{minipage}
}
\end{figure}

Figure \ref{motivation-perf} quantifies the performance of
hypervisor-manag\-ed die-stacked DRAM, and translation coherence's
impact on it. We normalize all performance numbers to the runtime of a
system with only off-chip DRAM and no high-bandwidth die-stacked DRAM
({\sf no-hbm}). Further, we show an unachievable best-case scenario
where all data fits in an infinite-sized die-stacked memory ({\sf
  inf-hbm}). After profiling several paging strategies (evaluated in
detail in Sec. \ref{results}), we plot the best-performing ones with
the {\sf curr-best} bars. These results assume cumbersome software
translation coherence mechanisms. In contrast, the {\sf achievable}
bars represent the potential performance of the best paging policies
with zero-overhead (and hence ideal) translation coherence.

Figure \ref{motivation-perf} shows that unachievable infinite
die-stacked DR\-AM can improve performance by 25-75\% ({\sf inf-hbm}
versus {\sf no-hbm}). Unfortunately, the current ``best'' paging
policies we achieve in KVM ({\sf curr-best}) fall far short of the
ideal {\sf inf-hbm} case. Translation coherence overheads are a big
culprit -- when these overheads are eliminated in {\sf achievable},
system performance comes within 3-10\% of the case with infinite
die-stacked DRAM capacity ({\sf inf-hbm}). In fact, Figure
\ref{motivation-perf} shows that translation coherence overheads can
be so high that they can prompt die-stacked DRAM to counterintuitvely
{\it worsen} performance. For example, {\sf data caching} and {\sf
  tunkrank} actually suffer 23\% and 10\% performance degradations in
{\sf curr-best}, respectively, despite using high-bandwi\-dth
die-stacked memory. Though omitted to save space, we have also
profiled the Xen hypervisors and found similar trends (presented in
Sec. \ref{results}). Overall, translation coherence overheads threaten
the use of die-stacked, and indeed any heterogeneous, memory.

\subsection{Page Remapping Anatomy}\label{page-remapping-anatomy}

\begin{figure}[t]
\centering
{
\begin{minipage}[t]{0.48\textwidth}
\centering
\vspace{-4mm}
\epsfig{file=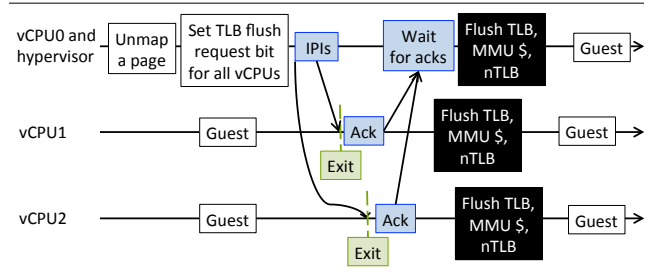, scale=0.34, angle=90}
\vspace{-38mm}
\caption{\small Sequence of operations associated with a page
  unmap. Initiator to target IPIs are shown in blue
  \textcircled{\small 1}, VM exits are shown in green
  \textcircled{\small 2}, and translation structure flushes are shown
  in black \textcircled{\small 3}. }
\label{kvm-unmap}
\end{minipage}
}
\end{figure}

We now shed light on the sources of overheads from translation
coherence. While we use page migration between off-chip and
die-stacked DRAM as our driving example, the same mechanisms are used
today to migrate pages between NUMA memories, or to defragment memory,
etc.

When a VM is configured, KVM assigns it virtual CPU threads or
vCPUs. Figure \ref{kvm-unmap} assumes 3 vCPUs executing on physical
CPUs. Suppose vCPU 0 frequently demands data in GVP 3, which maps to
GPP 8 and SPP 5, and that SPP 5 resides in off-chip DRAM. The
hypervisor may want to migrate SPP 5 to die-stacked memory (e.g., SPP
512) to improve performance. On a VM exit (assumed to have occurred
prior in time to Figure \ref{kvm-unmap}), the hypervisor modifies the
nested page table to update the SPP, triggering translation
coherence. There are three problems with this:

\vspace{2mm}\noindent{\bf All vCPUs are identified as targets:} Figure
\ref{kvm-unmap} shows that the hypervisor initiates translation
coherence by setting the TLB flush request bit in every vCPU's {\sf
  kvm\_vcpu} structure. {\sf kvm\_vcpu} stores vCPU state; when a vCPU
is scheduled on a physical CPU, it provides register content,
instruction pointers, etc. By setting these bits, the hypervisor
signals that TLB, MMU cache, and nTLB entries need to be flushed.

Ideally, we would like the hypervisor to identify only the CPUs that
actually cache the stale translation as targets. The hypervisor does
spare physical CPUs that never executed the VM. However, it flushes
all physical CPUs that ran any of the vCPUs of the VM, regardless of
whether they cache the modified page table entries. 

%problem in doing so, which we discuss in subsequent sections. As such,
%all vCPUs are conservatively identified as targets.

\begin{comment}
One might consider (similar to Linux) tracking the subset of physical
 Naturally, this may be
conservative since the translations may have been evicted from the
TLBs, etc. Unfortunately, however, hypervisors are unable to achieve
even this type of coarse-grained target identification. In a nutshell,
hypervisors (in full- rather than para-virtualized scenarios
\cite{adams:comparison}) cannot easily identify the subset of vCPUs
and physical CPUs that individual processes have executed on. Instead,
they conservatively identify {\it all} vCPUs as targets.
\end{comment}

\vspace{2mm}\noindent{\bf All vCPUs suffer VM exits:} In the next
step, the hypervisor launches inter-processor interrupts (IPIs) to all
the vCPUs. IPIs use the processor's advanced programmable interrupt
controllers (APICs). APIC implementations vary; depending on the APIC
technology, KVM converts broadcast IPIs into a loop of individual
IPIs, or a loop across processor clusters. We have profiled the
overheads of IPIs using microbenchmarks on Haswell systems, and like
past work \cite{oskin:sw, villavieja:didi}, find that they are
expensive, consuming thousands of clock cycles. If the receiving CPUs
are running vCPUs, they suffer VM exits, compromising
\textcircled{\small 3} from Sec. \ref{introduction}. Targets then
acknowledge the initiator, which is paused waiting for all vCPUs to
respond.

\vspace{2mm}\noindent{\bf All translation structures are flushed:} The
next step is to invalidate stale mappings in translation structure
entries. Current architectures provide ISA and microarchitectural
support for this via, for example, {\sf invlpg} instructions in x86,
etc. There are two caveats however. First, these instructions need the
GVP of the modified nested page table mapping to identify the TLB
entries that need to be invalidated. This is largely because modern
TLBs maintain GVP bits in the tag. While this is a good design choice
for non-virtualized systems, it is problematic for virtualized systems
because hypervisors do not have easy access to GVPs. Instead, they
have GPPs and SPPs. Consequently, KVM, Xen, etc., flush all TLB
contents when they modify a nested page table entry, rather than
selectively invalidating TLB entries. Second, there are currently no
instructions to selectively invalidate MMU caches or nTLBs, even
though they are tagged with GPPs and SPPs. The is because the marginal
benefits of adding ISA support for selective MMU cache and nTLB
invalidation are limited when the more performance-critical TLBs are
flushed.

\begin{comment}
One might consider solving this problem by re-designing the
guest-hypervisor interface. One possibility might be to create a
communication channel for the guest OS to pass information about GVP
changes to the hypervisor and guarantee synchronization of this
information between the guest and hypervisor. Unfortunately,
synchronizing on every guest page fault, address space switch, etc.,
is expensive, and re-introduces problems similar to those with shadow
paging \cite{ahn:revisiting}. In an alternative approach, we may
create a communication channel for the hypervisor to query GVP
information from the guest. However, the guest OS must not change GVPs
while the hypervisor uses it, introducing design complexity and
constraining guest OS operation. Overall, while software solutions
might be possible, they require complex changes to existing
guest-hypervisor interfaces.
\end{comment}

\subsection{Hardware Versus Software Solutions}

It is natural to ask whether translation coherence problems can be
solved with smarter software. We have studied this possibility and
have concluded that hardware solutions are superior. Fundamentally,
software solutions only partially solve the problem of flushing all
translation structures, and cannot solve the problem of identifying
all vCPUs as translation coherence targets and prompting VM exits.

Consider the problem of flushing all translation structures. One might
consider tackling this problem by modifying the guest-hypervisor
interface to enable the hypervisor to use existing ISA support (e.g.,
{\sf invlpg} instructions) to selectively invalidate TLB entries. But
this only fixes TLB invalidation -- no architectures today maintain
selective invalidation instructions for MMU caches and nTLBs, so these
would still have to be flushed.

Even if this problem could be solved, making target-side translation
coherence handling lightweight is challenging. Fundamentally, handling
translation coherence in software means that a context switch of the
CPUs is unavoidable. One alternative to expensive VM exits might be to
switch to lighterweight interrupts to query the guest OS for GVP-SPP
mappings. Unfortunately, even these interrupts remain
expensive. Specifically, we profiled interrupt costs using
microbe\-nchmarks on Intel's Haswell machines and found that they
require 640 cycles on average, which is just half of the average of
1300 cycles required for a VM exit. {\sf HATRIC}, however, entirely
eliminates these costs by {\it never} disrupting the operation of the
guest OS or requiring context switching.

\section{Hardware Design}\label{hardware-design}

We now detail {\sf HATRIC's} design, focusing mostly on
hyp\-ervisor-initiated paging which modifies the nested page
table. {\sf HATRIC} achieves all three goals set out in
Sec. \ref{introduction}. It does so by adding co-tags to translation
structures to achieve precise invalidation. It then exposes these
co-tags to the cache coherence protocol to precisely identify
coherence targets and to eliminate VM exits.

\subsection{Co-Tags}\label{co-tags}

\begin{figure}[t]
\centering {
\begin{minipage}[t]{0.48\textwidth}
\centering
\vspace{-4mm}
\epsfig{file=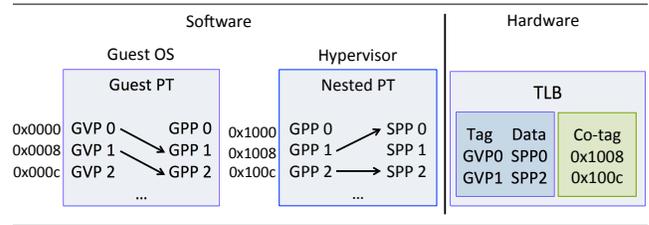, scale=0.34, angle=90}
\vspace{-44mm}
\caption{\small We add co-tags to store the system physical addresses
  where nested page table entries are stored. In our final
  implementation, we only store a subset of the system physical
  address bits.}
\vspace{-4mm}
\label{co-tags-figure}
\end{minipage}
}
\end{figure}

We describe co-tags by discussing what they are, what they accomplish,
how they are designed, and who sets them.

\vspace{2mm}\noindent{\bf What are co-tags?} Consider the page tables
of Figure \ref{co-tags-figure} and suppose that the hypervisor
modifies the GPP 2-SPP 2 nested page table mapping, making the TLB entry
caching information about SPP 2 stale. Since the TLB caches GVP-SPP
mappings rather than GPP-SPP mappings, this means that we'd like to
selectively invalidate GVP 1-SPP 2 from the TLB, and although not
shown, corresponding MMU cache and nTLB entries. Co-tags allow us to
do this by acting as tag extensions that allow precise identification
of translations when the hypervisor does not know the GVP. Co-tags
store the system physical address of the nested page table entry ({\sf
  nL1} from the bottom-most row in Figure
\ref{pt-translation-structures}). For example, GVP 1-SPP 2 uses the
nested page table entry at system physical address {\sf 0x100c}, which
is stored in the co-tag.

\vspace{2mm}\noindent{\bf What do co-tags accomplish?}  Co-tags not
only permit precise translation information identification but can
also be piggybacked on existing cache coherence protocols.  When the
hypervisor modifies a nested page table translation, cache coherence
protocols detect the modification to the system physical address of
the page table entry. Ordinarily, all private caches respond so that
only one amongst them holds the up-to-date copy of the cache line
storing the nested page table entry. With co-tags, {\sf HATRIC}
extends cache coherence as follows. Coherence messages, previously
restricted to just private caches, are now also relayed to translation
structures. Co-tags are used to identify which (if any) TLB, MMU
cache, and nTLB entries correspond to the modified nested page table
cache line. Overall, this means that co-tags: \textcircled{\small a}
pick up on nested page table changes entirely in hardware, without the
need for IPIs, VM exits, or {\sf invlpg} instructions;
\textcircled{\small b} rely on, without fundamentally changing,
existing cache coherence protocols; \textcircled{\small c} permit
selective TLBs, MMU caches, and nTLBs rather than flushes.

\vspace{2mm}\noindent{\bf How are co-tags implemented?} Co-tags have
one important drawback. System physical addresses on 64-bit systems
require 8 bytes. If all 8 bytes are realized in the co-tag, each TLB
entry doubles in size. MMU cache and nTLB entries triple in
size. Since address translation can account for 13-15\% of processor
energy \cite{fan:energy, karakostas:energy, juan:reducing,
  kadayif:compiler,sodani:race}, these area and associated energy
overheads are unacceptable.

Therefore, we decrease the resolution of co-tags, using fewer
bits. This means that groups, rather than individual TLB entries may
be invalidated when one nested page table entry is changed. However,
judiciously-sized co-tags generally achieve a good balance between
invalidation precision, and area/energy overheads. Sec. \ref{results}
shows, using detailed RTL modeling, that 2-byte co-tags (a per-core
area overhead of 2\%) strike a good balance. We specify the exact
subset of address bits make up the co-tag in subsequent sections.

\vspace{2mm}\noindent{\bf Who sets co-tags?} For good performance,
co-tags must be set by hardware without an OS or hypervisor
interrupt. {\sf HATRIC} uses the page table walker to do this. On TLB,
MMU cache, and nTLB misses, the page table walker performs a
two-dimensional page table walk. In so doing, it infers the system
physical address of the page table entries and stores it in the TLB,
MMU cache, and nTLB co-tags.

\subsection{Integration with Cache Coherence}\label{cache-coherence} 

Modern cache coherence protocols can integrate not only readable and
writable private caches, but also read-only instruction caches (though
instruction caches do not have to be read-only). Since TLBs, MMU
caches, and nested TLBs are fundamentally read-only structures, {\sf
  HATRIC} integrates them into the existing cache coherence protocol
in a manner similar to read-only instruction caches. Beyond this, {\sf
  HATRIC} has minimal impact on the cache coherence protocol. We
describe {\sf HATRIC's} operation on a directory-based MESI protocol,
with the coherence directories located at the shared LLC cache
banks. Without loss of generality, we use dual-grain coherence
directories from recent work \cite{zebchuk:multi}.

\vspace{2mm}\noindent{\bf Translation structure coherence states:}
Since translation structures are read-only, their entries require only
two coherence states: Shared (S), and Invalid (I). These two states
may be realized using per-entry {\sf valid} bits. When a translation
is entered into the TLB, MMU cache, or nTLB, the {\sf valid} bit is
set, representing the S state; the translation can be accessed by the
local CPU. The translation structure entry remains in this state until
it receives a coherence message. Co-tags are compared to incoming
messages; when an invalidation request matches the co-tag, the
translation entry is invalidated.

\begin{figure}[t]
\centering
{
\begin{minipage}[t]{0.48\textwidth}
\centering
\vspace{-4mm}
\epsfig{file=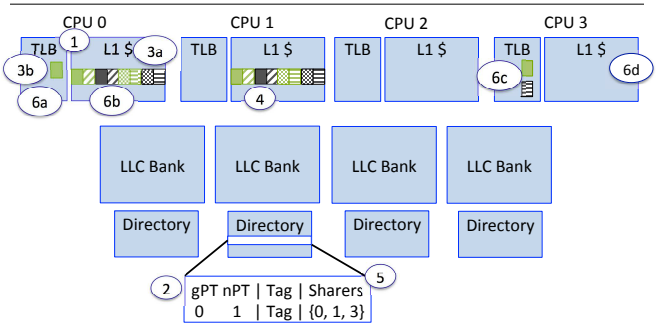, scale=0.34, angle=90}
\vspace{-28mm}
\caption{\small Coherence directories identify translation structures
  caching page table entries, aside from private L1 cache contents.}
\vspace{-4mm}
\label{coherence-1}
\end{minipage}
}
\end{figure}

\vspace{2mm}\noindent{\bf Translation coherence initiators:} Consider
Figure \ref{coherence-1}. Before detailing the numbered transactions,
let us consider {\sf HATRIC's} components. We show a 4-CPU system,
with private L1 caches, 4 shared LLC banks, and per-bank coherence
directories. We show TLBs and though they also exist, we omit MMU
caches and nTLBs to save space. MMU caches and nTLBs interact with the
cache coherence protocol in a manner that mirrors TLBs. We show 8
cached page table entries, represented as green and black boxes.
Translation coherence is initiated by the hardware page table walker
or OS/hypervisor software.

\vspace{2mm}\noindent {\it Page table walkers}: These are hardware
finite state machines that are invoked on TLB misses. Walkers traverse
the page tables and are responsible for filling translation
information into the translation structures and setting the
co-tags. Walkers cannot map or unmap pages.

\vspace{2mm}\noindent {\it OS and hypervisor:} These can traverse,
map, and unmap page table entries using standard load/store
instructions. {\sf HATRIC} picks up these changes, and keeps all
private cache and translation structures coherent.

\vspace{2mm}\noindent{\bf Coherence directory:} {\sf HATRIC} minimally
changes the coherence directory. Key design considerations are:

\vspace{2mm} \noindent{\it Directory entry changes:} Figure
\ref{coherence-1} shows that the coherence directory tracks non-page
table and page table cache lines. We make a minor change to directory
entries, adding two bits to record whether cache lines belong to a
guest page table ({\sf gPT}) or nested page table ({\sf nPT}). {\sf
  HATRIC} uses these bits to identify the case when a line holding
page table data is modified in the private caches. When this happens,
coherence transactions need to be sent to the translation structures.

The {\sf nPT} and {\sf gPT} bits are set by the hardware page table
walkers on fills to the TLBs, MMU caches, and nTLBs. One might
initially expect this to be problematic in the case where the OS or
hypervisor reads or writes a page table cache line in software. In
reality however, this does not present correctness issues. Two
situations are possible. In the first situation, the page table walker
has previously accessed the cache line, and has already set the {\sf
  nPT} or {\sf gPT} bit in the cache line's directory entry. There are
no correctness issues in this case. In the second situation, the OS or
hypervisor reads or writes a page table cache line that has previously
never been looked up by the page table walker. In this case, there is
actually no need to set the {\sf nPT} or {\sf gPT} bits in the
coherence directory entry yet since no translations from this line are
cached in the TLB, MMU cache, or nTLB anyway. Modifying the cache line
at this point does not require coherence messages to be sent to the
translation structures. When the page table walker does eventually
access a translation from this cache line and fills it into the
translation structures, it checks the {\sf access} bit already
maintained by x86-64 translation entries. The {\sf access} bit records
whether an entry has previously been filled into the TLB or accessed
by the page table walker \cite{lustig:coatcheck}. If this bit is
clear, this means that the entry (and hence the cache line it resides
in) has not been accessed by the page table walker yet. In this case,
the page table walker sends a message to the coherence directory to
update the {\sf nPT} and {\sf gPT} bits of the relevant cache line.

\vspace{2mm}\noindent{\it Coherence granularity:} Figure
\ref{coherence-1} shows that directory entries store information at
the cache line granularity. x86-64 systems cache 8 page table entries
per 64-byte cache line. Hence, similar to false sharing in caches
\cite{luo:laser}, {\sf HATRIC} conservatively invalidates all
translation structure entries caching these 8 page table entries, even
if only a single page table entry is modified. For example, consider
CPU 3 in Figure \ref{coherence-1}, where the TLB caches two
translations mapped to the same cache line. If any CPU modifies either
one of these translations, {\sf HATRIC} has to invalidate both TLB
entries. This has implications on the size of co-tags. Recall that in
Sec. \ref{co-tags}, we stated that co-tags use a subset of the address
bits. We want use the least significant, and hence, highest entropy
bits as co-tags. But since cache coherence protocols track groups of
8 translations, co-tags do not store the 3 least significant address
bits. Our 2 byte co-tags use bits 19-3 of the system physical address
storing the page table. Naturally, this means that translations from
different addresses in the page table may alias to the same co-tag. In
practice, this has little adverse affect on {\sf HATRIC's}
performance.

\vspace{2mm}\noindent {\it Coherence specificity issues:} To simplify
hardware, coherence directories do not track where among the private
caches, TLB, MMU cache, and nTLB the page table entries are
cach\-ed. Instead, coherence directories are {\it
  pseudo-specific}. For example, Figure \ref{coherence-1} shows that
CPU 0 caches page table entries in the TLB and L1 cache, CPU 1 only
caches them in the L1 cache, while CPU 3 only caches them in the
TLB. Nevertheless, the coherence directory's sharer list does not
capture this distinction. Therefore, when a CPU modifies page table
contents and invalidation messages need to be sent to the sharers,
they are relayed to the L1 caches {\it and} all translation
structures, regardless of which ones actually cache page tables. This
results in spurious coherence activity (e.g., CPU 3's L1 cache need not
be relayed an invalidation message for any of the page table entries
shown). In practice though, because modifications of the page table
are rare compared to other coherence activity, this additional traffic
is tolerable. Ultimately, the gains from eliminating high-latency
software TLB coherence far outweigh these relatively minor overheads
(see Sec. \ref{results}).

\vspace{2mm}\noindent {\it Cache and translation structure evictions:}
Directories track translations in a coarse-grained and pseudo-specific
manner. This has important implications on cache line
evictions. Ordinarily, when a private cache line is evicted, the
coherence directory is relayed a message to update the line's sharer
list \cite{zebchuk:multi}. An up-to-date sharer list eliminates
spurious coherence traffic to this line in the future. We continue to
employ this strategy for non-page table cache lines but use a slightly
different approach for page tables. When a cache line holding page
table entries is evicted, its content may still be cached in the TLB,
MMU cache, nTLB. Even worse, other translations with matching co-tags
may still be residing in the translation structures. One option may be
to detect all translations with matching co-tags and invalidate
them. This hurts energy because of the additional translation
structure lookups, and performance because of unnecessary TLB, MMU
cache, and nTLB entry invalidations.

\begin{figure}[t]
\centering
{
\begin{minipage}[t]{0.48\textwidth}
\centering
\vspace{-4mm}
\epsfig{file=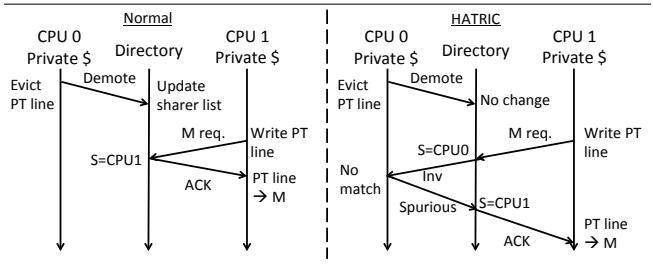, scale=0.34, angle=90}
\vspace{-38mm}
\caption{\small Coherence activity from the eviction of a cache line
  holding page table entries from CPU0's private cache. {\sf HATRIC}
  updates sharer list information lazily in response to cache line
  evictions.}
\label{lazy}
\end{minipage}
}
\end{figure}

Figure \ref{lazy} shows how {\sf HATRIC} handles this problem,
contrasting it with traditional cache coherence. Suppose CPU 0 evicts
a cache line with page table entries. Both approaches relay a message
to the coherence directory. Ordinarily, we remove CPU 0 from the
sharer list. However, if {\sf HATRIC} sees that this message
corresponds to a cache line storing a page table (by checking the
directory entry's page table bits), the sharer list is untouched.
This means that if CPU 1 subsequently writes to the same cache line,
{\sf HATRIC} sends spurious invalidate messages to CPU 0, unlike
traditional cache coherence. However, we mitigate frequency of
spurious messages; when CPU 0 sees spurious coherence traffic, it
sends a message back to the directory to demote CPU 0 from the sharer
list. Sharer lists are hence lazily updated. For similar reasons,
evictions from translation structures also lazily update coherence
directory sharer lists.

\vspace{2mm}\noindent {\it Directory evictions:} Finally, past work
shows that coherence directory entry evictions require
back-invalidations of the associated cache lines in the cores
\cite{zebchuk:multi}. This is necessary for correctness; all lines in
private caches must always have a directory entry. {\sf HATRIC}
extends this approach to relay back-invalidations to the TLBs, MMU
caches, and nTLBs too.

\subsection{Putting It All Together}\label{putting-it-all-together}

Figure \ref{coherence-1} details {\sf HATRIC's} overall
operation. Initially, CPU 0's TLB and L1 caches are empty. On a memory
access, CPU 0 misses in the TLB and walks the page table
\textcircled{\small 1}. Whenever a request is satisfied from a page
table line in the L1 cache in the M, E, or S state, there is no need
to initiate coherence transactions. However, suppose that the last
memory reference in the page table walk from Figure
\ref{pt-translation-structures} is absent in the L1 cache. A read
request is sent to the coherence directory in step \textcircled{\small
  2}.

Two scenarios are possible. In the first, the translation may be
uncached in the private caches, and there is no coherence directory
entry. A directory entry is allocated and the {\sf gPT} or {\sf nPT}
bit is set. In the second scenario (shown in Figure
\ref{coherence-1}), the request matches an existing directory
entry. The {\sf nPT} bit already is set and {\sf HATRIC} reads the
sharer list which identifies CPUs 1 and 3 as also caching the desired
translation (and the 7 adjacent translations in the cache line) in
shared state. In response, the cache line with the desired
translations is sent back to CPU 0 (from CPU 1, 3, or memory,
whichever is quicker), updating the L1 cache \encircle{\small 3a} and
TLB \encircle{\small 3b}. Subsequently, the sharer list adds CPU 0.

Now suppose that CPU 1 runs the hypervisor and unmaps the solid green
translation from the nested page table in step \textcircled{\small
  4}. To transition the L1 cache line into the M state, the cache
coherence protocol relays a message to the coherence directory. The
corresponding directory entry is identified in \textcircled{\small 5},
and we find that CPU 0 and 3 need to be sent invalidation
requests. However, the sharer list is (i) coarse-grained and (ii)
pseudo-specific. Because of (i), CPU 0 has to invalidate not only its
TLB entry \encircle{\small 6a} but also 8 translations in the L1 cache
\encircle{\small 6b}, and CPU 3 has to invalidate the 2 TLB entries
with matching co-tags \encircle{\small 6c}. Because of (ii), CPU 1's L1
cache receives a spurious invalidation message \encircle{\small 6d}.

\begin{comment}

Steps \textcircled{\small 6a}-\textcircled{\small 6d}
constitute these messages. Because page table entries are tracked at a
cache line granularity by the coherence protocol, CPU0 has to
invalidate 8 translations, as per usual \textcircled{\small 6b}. {\sf
  HATRIC} uses co-tags to go beyond traditional cache coherence to
invalidate the TLB's copy of the solid green translation too
\textcircled{\small 6a}. CPU3 also attempts to invalidate its copies
of the now stale translation. However, because of coarser-grained
co-tags, CPU3 has to invalidate the striped black translation in
addition to the solid green one \textcircled{\small 6c}. Further, as
the sharer list does not more precisely indicate where within the
translation structures and L1 caches a page table entry resides,
CPU3's L1 cache is spuriously looked up \textcircled{\small 6d}.
\end{comment}

%\begin{figure}[t]
%\centering
%{
%\begin{minipage}[t]{0.48\textwidth}
%\centering
%\vspace{-4mm}
%\epsfig{file=coherence-2.ps, scale=0.34, angle=90}
%\vspace{-28mm}
%\caption{\small Coherence directories lazily update information about
%  the cores sharing a cache line, increasing on-chip network traffic.}
%\label{coherence-2}
%\end{minipage}
%}
%\end{figure}

\subsection{Other Key Observations}\label{other-key-observations}

\noindent{\bf Scope:} {\sf HATRIC} is applicable to virtualized and
non-virtualized systems. For the latter, the co-tags may simply be
used to store the physical addresses of page tables. Further, while we
have focused on nested page table coherence, {\sf HATRIC} can also be
trivially modified to support shadow page tables too
\cite{ahn:revisiting}. The co-tags merely have to store the memory
addresses where shadow page tables are stored.

\vspace{2mm}\noindent{\bf Metadata updates:} Beyond software changes
to the translations, they may also be changed by hardware page table
walkers. Specifically, page table walkers update dirty and access bits
to aid page replacement policies \cite{pham:glue}. But since these
updates are picked up by the standard cache coherence protocol, {\sf
  HATRIC} naturally handles these updates too.

\vspace{2mm}\noindent{\bf Prefetching optimizations:} Beyond simply
invalidating stale translation structure entries, {\sf HATRIC} could
potentially directly update (or prefetch) the updated mappings into
the translation structures. Since a thorough treatment of these
studies requires an understanding of how to manage translation access
bits while speculatively prefetching into translation structures
\cite{lustig:coatcheck}, we leave this for future work.

\vspace{2mm}\noindent{\bf Coherence protocols:} We have studied a MESI
directory based coherence protocol but we have also implemented {\sf
  HATRIC} atop MOESI protocols too, as well as snooping protocols like
MESIF \cite{goodman:mesif}. {\sf HATRIC} requires no fundamental
changes to support these protocols.

\vspace{2mm}\noindent{\bf Synonyms and superpages:} {\sf HATRIC}
naturally handles synonyms or virtual address aliases. This is because
synonyms are defined by unique translations in separate page table
locations, and hence separate system physical addresses. Therefore,
changing or removing a translation has no impact on other translations
in the synonym set, allowing {\sf HATRIC} to be agnostic to
synonyms. Similarly, {\sf HATRIC} supports superpages, which also
occupy unique translation entries and can hence be easily detected
by co-tags.

\vspace{2mm}\noindent{\bf Multiprogrammed workloads:} One might expect
that when an application's physical page is remapped, there is no need
for translation coherence activities to the other applications,
because they operate on distinct address spaces. Unfortunately,
however, hypervisors do not know which physical CPUs an application
executed on; all they know is the vCPUs and the physical the entire VM
uses. Therefore, the hypervisor conservatively flushes the even the
translation structures of CPUs that never ran the offending
application.  {\sf HATRIC} completely eliminates this problem by
precisely tracking the correspondence between translations and CPUs.

\vspace{2mm}\noindent{\bf Comparison to past approaches:} {\sf HATRIC}
is inspired by past work on {\sf UNITD} \cite{romanescu:unified}. Like
{\sf HATRIC}, {\sf UNITD} piggybacks translation coherence atop cache
coherence protocols. Unlike {\sf HATRIC} however, {\sf UNITD} cannot
support virtualized systems or MMU cache and nTLB coherence. Further,
{\sf HATRIC} uses energy-frugal co-tags instead of {\sf UNI\-TD's}
large reverse-lookup CAM circuitry, achieving far grea\-ter energy
efficiency. We showcase this in Sec. \ref{results} where we compare
the efficiency of {\sf HATRIC} versus an enhanced {\sf UNITD} design
for virtualization. Beyond {\sf UNITD}, past work on {\sf DiDi}
\cite{villavieja:didi} also targets translation coherence for
non-virt\-ualized systems. Similarly, recent work investigates
translation coherence overheads in the context of die-stacked DRA\-M
\cite{oskin:sw}. While this work mitigates translation coherence
overheads, it does so specifically for non-virtualized x86
architectures, and ignores MMU caches and nTLBs. Finally, recent work
uses software mechanisms to reduce translation overheads for guest
page table modifications \cite{ouyang:shoot4u}, while {\sf HATRIC}
also solves the problem of nested page table coherence.

\section{Methodology}\label{methodology}
Our experimental methodology has two steps. First, we modify KVM to
implement paging on a two-level memory with die-stacked DRAM. Second,
we use detailed cycle-accurate simulation to assess performance and
energy.

\subsection{Die-Stacked DRAM Simulation}
We evaluate {\sf HATRIC's} performance on a detailed cycle-accurate
simulation framework that models the operation of a 32-CPU Haswell
processor. We assume 2GB of die-stacked DRAM with 4$\times$ the
bandwidth of slower 8GB off-chip DRA\-M, similar to prior work
\cite{oskin:sw}. Each CPU maintains 32KB L1 caches, 256KB L2 caches,
64-entry L1 TLBs, 512-entry L2 TLBs, 32-entry nTLBs
\cite{bhargava:asplos}, and 48-entry paging structure MMU caches
\cite{bhattacharjee:mmu}. Further, we assume a 20MB LLC.  We model the
energy usage of this system using the CACTI framework
\cite{muralimanohar:cacti}. We use Ubuntu 15.10 Linux as our guest
OS. Further, we evaluate {\sf HATRIC} in detail using KVM. Beyond
this, we have also run Xen to highlight {\sf HATRIC's} generality with
other hypervisors.

We use a trace-based approach to drive our simulation framework. We
collect instruction traces from our modified hypervisors with 50
billion memory references using a modified version of Pin which tracks
all GVPs, GPPs, and SPPs, as well as changes to the guest and nested
page tables. In order to collect accurate paging activity, we collect
these traces on a real-system. Ideally, we would like this system to
use die-stacked DRAM but since this technology is in its infancy, we
are inspired by recent work \cite{oskin:sw} to modify a real-system to
mimic the activity of die-stacking. We take an existing multi-socket
NUMA platform, and by introducing contention, creates two different
speeds of DRAM. We use a 2-socket Intel Xeon E5-2450 system, running
our software stack. We dedicate the first socket for execution of the
software stack and mimicry of fast or die-stacked DRAM. The second
socket mimics the slow or off-chip DRAM. It does so by running several
instances of {\sf memhog} on its cores. Similar to prior work
\cite{pham:CoLT, pham:glue, bhattacharjee:translation}, we use {\sf
  memhog} to carefully generate memory contention to achieve the
desired bandwidth differential between the fast and slow DRAM of
4$\times$. By using Pin to track KVM and Linux paging code on this
infrastructure, we accurately generate instruction traces to test {\sf
  HATRIC}.

\subsection{KVM Paging Policies}\label{kvm-paging-policies}

Our goal is to showcase the overheads imposed by translation coherence
on paging decisions rather than design the optimal paging policy,
leaving this for future work. So, we pick well-known paging policies
that cover a wide range of design options. For example, we have studied
FIFO and LRU replacement policies, finding the latter to perform
better, as expected. We implement LRU policies in KVM by repurposing
Linux's well-known pseudo-LRU {\sf CLOCK} policy \cite{easton:use}. LRU
alone doesn't always provide good performance since it is expensive to
traverse page lists to identify good candidates for eviction from
die-stacked memory. Instead, performance is improved by moving this
operation off the critical path of execution; we therefor
pre-emptively evict pages from die-stacked memory so that a pool of
free pages are always maintained. We call this {\it migration daemon}
and combine it with LRU. We have also investigated the benefits of
page prefetching; that is, when an application demand fetches a page
from off-chip to die-stacked memory, we also prefetch a set number of
adjacent pages. Generally, we have found that the best paging policy
uses a combination of these approaches.

\subsection{Workloads}\label{workloads}

Our focus is on two sets of workloads. The first set comprises
applications that benefit from the higher bandwidth of die-stacked
memory. We use {\sf canneal} and {\sf facesim} from the Parsec suite
\cite{bienia:parsec}, {\sf data caching} and {\sf tunkrank} from
Cloudsuite \cite{falsafi:cloudsuite}, and {\sf graph500} as part of
this group. We also create 80 multiprogrammed combinations of
workloads from all the Spec applications to showcase the problem of
imprecise target identification in virtualized translation coherence.

Our second group of workloads is made up of smaller-footprint
applications whose data largely fits within the die-stacked DRAM. We
use these workloads to evaluate {\sf HATRIC's} overheads in situations
where hypervisor-mediated paging (and hence translation coherence)
between die-stacked and off-chip DRAM is rarer. We use the remaining
Parsec applications, and Spec applications for these studies.

\section{Evaluation}\label{results}

\begin{figure}[t]
\centering
{
\begin{minipage}[t]{0.48\textwidth}
\centering
\vspace{-4mm}
\epsfig{file=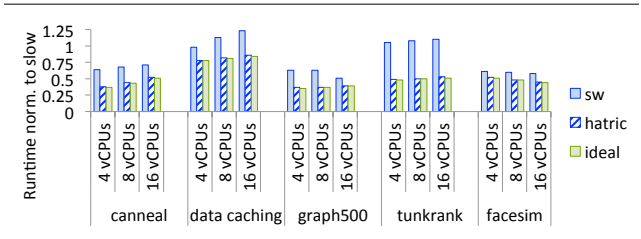, scale=0.34, angle=90}
\vspace{-40mm}
\caption{\small For varying vCPUs, runtime of the best KVM paging
  policy without {\sf HATRIC} ({\sf sw}), with {\sf HATRIC} ({\sf
    hatric}), and with zero-overhead translation coherence ({\sf
    ideal}). All results are normalized to the case without
  die-stacked DRAM.}
\label{perf-1}
\end{minipage}
}
\end{figure}

\vspace{2mm} \noindent {\bf Performance as a function of vCPU counts:}
Figure \ref{perf-1} shows {\sf HATRIC's} runtime, normalized as a
fraction of application runtime in the absence of any die-stacked
memory ({\sf no-hbm} from Figure \ref{motivation-perf}). We compare
runtimes for the best KVM paging policies ({\sf sw}), {\sf HATRIC},
and ideal unachievable zero-overhead translation coherence ({\sf
  ideal}). Further, we vary the number of vCPUs per VM and observe the
following.

{\sf HATRIC} is {\it always} within 2-4\% of the {\sf ideal}
performance. In some cases, {\sf HATRIC} is instrumental in achieving
any gains from die-stacked memory at all. Consider {\sf data caching},
which slows down when using die-stacked memory, because of translation
coherence overheads. {\sf HATRIC} cuts runtimes down to roughly 75\%
of the baseline runtime in all cases.

Figure \ref{perf-1} also shows that {\sf HATRIC} is valuable at all
vCPU counts. In some cases, more vCPUs exacerbate translation
coherence overheads. This is because IPI broadcasts become more
expensive and more vCPUs suffer VM exits.  This is why, for example,
{\sf data caching} and {\sf tunkrank} become slower (see {\sf sw})
when vCPUs increase from 4 to 8. {\sf HATRIC} eliminates these
problems, flattening runtime improvements across vCPU counts. In other
scenarios, fewer vCPUs worsen performance since each vCPU performs
more of the application's total work. Here, the impact of a full TLB,
nTLB, and MMU cache flush for every page remapping is very expensive
(e.g., {\sf graph500} and {\sf facesim}). Here, {\sf HATRIC} again
eliminates these overheads almost entirely.

\begin{figure}[t]
\centering
{
\begin{minipage}[t]{0.48\textwidth}
\centering
\vspace{-4mm}
\epsfig{file=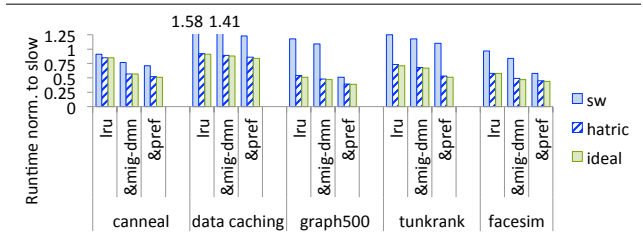, scale=0.34, angle=90}
\vspace{-40mm}
\caption{\small {\sf HATRIC's} performance benefits as a function of
  KVM paging policies, with LRU, migration daemons ({\sf mig-dmn}),
  and prefetching ({\sf pref.}). Results are normalized to the case
  without die-stacked DRAM.}
\label{perf-2}
\end{minipage}
}
\end{figure}

\vspace{2mm} \noindent {\bf Performance as a function of paging
  policy:} Figure \ref{perf-2} also shows {\sf HATRIC} performance,
but this time as a function of different KVM paging policies. We study
three policies with 16 vCPUs. First, we show {\sf lru}, which
determines which pages to evict from die-stacked DRAM. We then add the
migration daemon ({\sf \&mig-dmn}), and page prefetching
({\sf\&pref}).

Figure \ref{perf-2} shows {\sf HATRIC} improves runtime substantially
for any paging policy. Performance is best when all techniques are
combined, but {\sf HATRIC} achieves 10-30\% performance improvements
even for just {\sf lru}. Furthermore, Figure \ref{perf-2} shows that
translation coherence overheads can often be so high that the paging
policy itself makes little difference to performance. Consider {\sf
  tunkrank}, where the difference between {\sf lru} versus the {\sf
  \&pref} bars is barely 2-3\%. With {\sf HATRIC}, however, paging
optimizations like prefetching and migration daemons help. %We believe
%that as the systems community begins to investigate hypervisor
%management of heterogeneous memories, {\sf HATRIC} will be crucial to
%any page remapping policy.

\begin{figure}[t]
\centering
{
\begin{minipage}[t]{0.48\textwidth}
\centering
\vspace{-4mm}
\epsfig{file=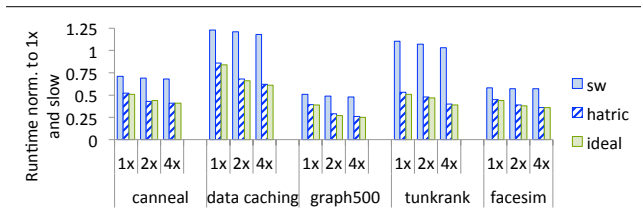, scale=0.34, angle=90}
\vspace{-42mm}
\caption{\small {\sf HATRIC's} performance benefits as a function of
  translation structure size. {\sf 1$\times$} indicates default sizes,
  {\sf 2$\times$} doubles sizes, and so on. All results are normalized
  to the case without die-stacked DRAM.}
\label{perf-3}
\vspace{-4mm}
\end{minipage}
}
\end{figure}

\vspace{2mm}\noindent{\bf Impact of translation structure sizes:} One
of {\sf HATRIC's} advantages is that it converts translation structure
flushes to selective invalidations. This improves TLB, MMU cache, and
nTLB hit rates substantially, obviating the need for expensive
two-dimensional page table walks. We expect {\sf HATRIC} to improve
performance even more as translation structures become bigger (and
flushes needlessly evict more entries). Figure \ref{perf-3} quantifies
the relationship. We vary TLB, nTLB, and MMU cache sizes from the
default (see Sec. \ref{methodology}) to double ({\sf 2$\times$}) and
quadruple ({\sf 4$\times$}) the number of entries.

Figure \ref{perf-3} shows that translation structure flushes largely
counteract the benefits of greater size. Specifically, the {\sf sw}
results see barely any improvement, even when sizes are
quadrupled. Inter-DRAM page migrations essentially flush the
translation structures so often that additional entries are not
effectively leveraged. Figure \ref{perf-3} shows that this is a wasted
opportunity since zero-overhead translation coherence ({\sf ideal})
actually does enjoy 5-7\% performance benefits. {\sf HATRIC} solves
this problem, comprehensively achieving within 1\% of the {\sf ideal},
thereby exploiting larger translation structures.

\begin{figure}[t]
\centering
{
\begin{minipage}[t]{0.48\textwidth}
\centering
\vspace{-4mm}
\epsfig{file=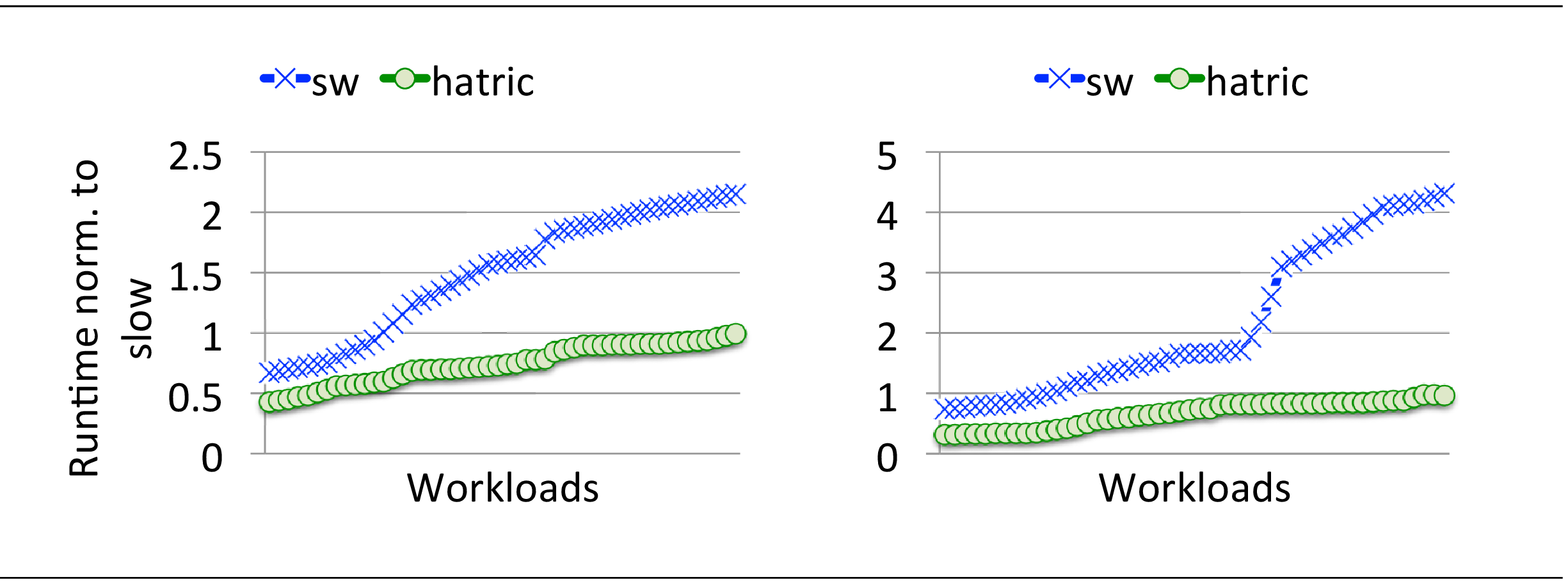, scale=0.34, angle=90}
\vspace{-42mm}
\caption{\small (Left) Weighted runtime for all 80 multi-programmed
  workloads on VMs without ({\sf actual}) and with {\sf HATRIC} ({\sf
    hatric}); (right) the same for the slowest application in mix.}
\label{multi-perf}
\vspace{-4mm}
\end{minipage}
}
\end{figure}

\vspace{2mm}\noindent{\bf Multi-programmed workloads:} We now focus on
multiprogrammed workloads made up sequential applications. Each
workload runs 16 Spec benchmarks on a Linux VM atop KVM. As is
standard for multiprogrammed workloads, we use two performance metrics
\cite{mutlu:bliss, mutlu:blacklisting}. The first is weighted runtime
improvement, which captures overall system performance. The second is
the runtime improvement of the slowest application in the workload,
capturing fairness.

Figure \ref{multi-perf} shows our results. The graph on the left plots the
weighted runtime improvement, normalized to cases without die-stacked
DRAM. As usual, {\sf sw} represents the best KVM paging policy. The
x-axis represents the workloads, arranged in ascending order of
runtime. The lower the runtime, the better the performance. Similarly,
the graph on the right of Figure \ref{perf-3} shows shows the runtime
of the slowest application in the workload mix; again, lower runtimes
indicate a speedup in the slowest application.

Figure \ref{multi-perf} shows that translation coherence can be disastrous
to the performance of multiprogrammed workloads. More than 70\% of the
workload combinations suffer performance degradation with
die-stacking. These applications suffer from unnecessary translation
structure flushes and VM exits, caused by software translation
coherence's imprecise target identification. Runtime is more than
2$\times$ for 11 workloads. Additionally, translation coherence
degrades application fairness. For example, in more than half the
workloads, the slowest application's runtime is (2$\times$)+ with a
maximum of (4$\times$)+. Applications that struggle are usually those
with limited memory-level parallelism that benefit little from the
higher bandwidth of die-stacked memory and instead, suffer from the
additional translation coherence overheads.

{\sf HATRIC} solves all these issues, achieving improvements for every
single weighted runtime, and even for each of the slowest
applications. In fact, {\sf HATRIC} entirely eliminating translation
coherence overheads, reducing runtime to 50-80\% of the baseline
without die-stacked DRAM. The key enabler is {\sf HATRIC's} precise
identification of coherence targets -- applications that do not need
to participate in translation coherence operations have their
translation structure contents left unflushed and do not suffer VM
exits.

\begin{figure}[t]
\centering {
\begin{minipage}[t]{0.48\textwidth}
\centering
\vspace{-4mm}
\epsfig{file=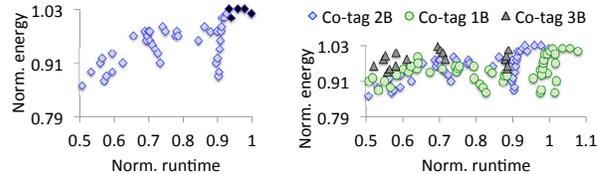, scale=0.34, angle=90}
\vspace{-44mm}
\caption{\small (Left) Performance-energy plots for default {\sf
    HATRIC} configuration compared to a baseline with the best paging
  policy; and (right) impact of co-tag size on performance-energy
  tradeoffs.}
\label{energy}
\vspace{-4mm}
\end{minipage}
}
\end{figure}

\vspace{2mm}\noindent{\bf Performance-energy tradeoffs:} Intuitively,
we expect that since {\sf HATRIC} reduces runtime substantially, it
should reduce static energy sufficiently to offset the higher energy
consumption from the introduction of co-tags. Indeed, this is true for
workloads that have sufficiently large memory footprints to trigger
inter-memory paging. However, we also assess {\sf HATRIC's} energy
implications on workloads that do not frequently remap pages (i.e.,
their memory footprints fit comfortably within die-stacked
DRAM). 

The graph on the left of Figure \ref{energy} plots all the workloads
including the single-threaded and multithreaded ones that benefit from
die-stacking and those whose memory needs fit entirely in die-stacked
DRAM. The x-axis plots the workload runtime, as a fraction of the
runtime of {\sf sw} results. The y-axis plots energy, similarly
normalized. We desire points that converge towards the lower-left
corner of the graph.

The graph on the left of Figure \ref{energy} shows that {\sf HATRIC}
always boosts performance, and almost always improves energy
too. Energy savings of 1-10\% are routine. In fact, {\sf HATRIC} even
improves the performance and energy of many workloads that do not page
between the two memory levels. This is because these workloads still
remap pages to defragment memory (to support superpages) and {\sf
  HATRIC} mitigates the associated translation coherence
overheads. There are some rare instances (highlighted in black) where
energy does exceed the baseline by 1-1.5\%. These are workloads for
whom efficient translation coherence does not make up for the
additional energy of the co-tags. Nevertheless, these overheads are
low, and their instances rare.

\vspace{2mm}\noindent{\bf Co-tag sizing:} We now turn to co-tag
sizing. Excessively large co-tags consume significant lookup and
static energy, while small ones force {\sf HATRIC} to invalidate too
many translation structures on a page remap. The graph on the right of
Figure \ref{energy} shows the performance-energy implications of
varying co-tag size from 1 to 3 bytes.

First and foremost, 2B co-tags -- our design choice -- provides the
best balance of performance and energy. While 3B co-tags track page
table entries at a finer granularity, they only modestly improve
performance over 2B co-tags, but consume much more energy. Meanwhile
1B co-tags suffer in terms of both performance and energy. Since 1B
co-tags have a coarser tracking granularity, they invalidate more
translation entries from TLBs, MMU caches, and nTLBs than larger
co-tags. And while the smaller co-tags do consume less lookup and
static energy, these additional invalidations lead to more expensive
two-dimensional page table walks and a longer system runtime. The end
result is an increase in energy too. 

\begin{figure}[t]
\centering
{
\begin{minipage}[t]{0.48\textwidth}
\centering
\vspace{-4mm}
\epsfig{file=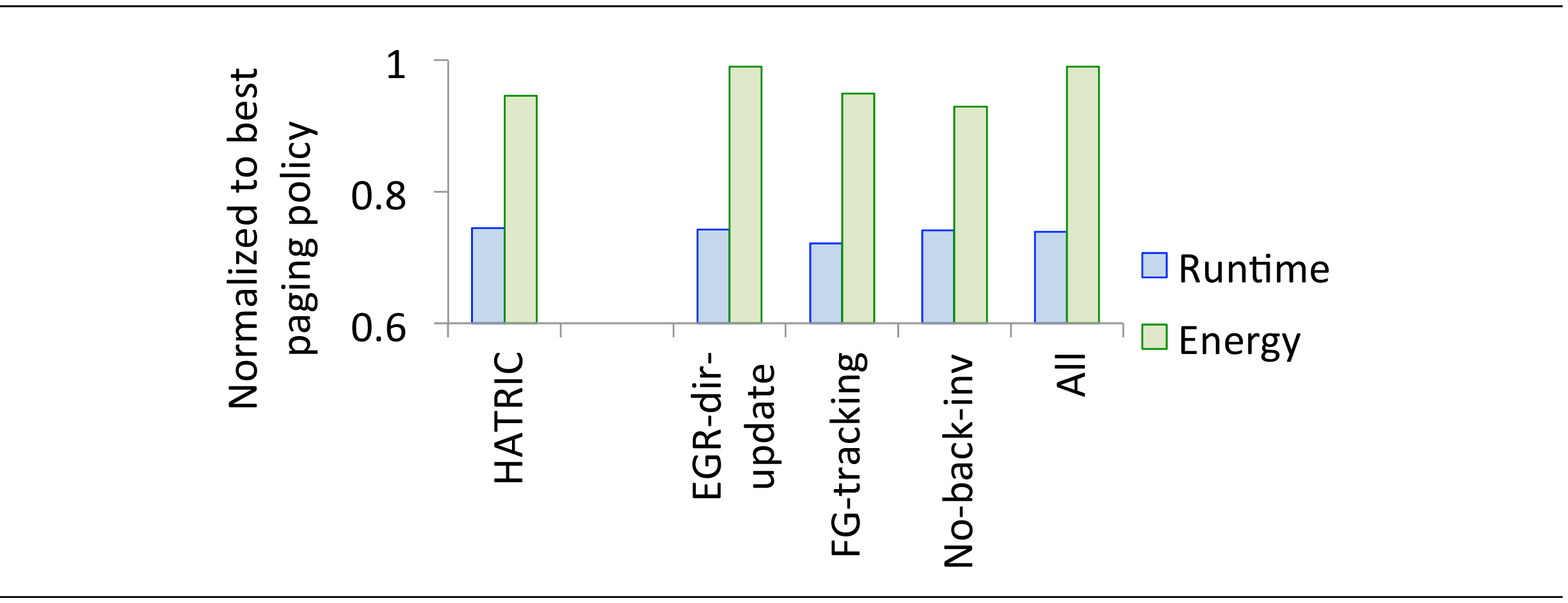, scale=0.34, angle=90}
\vspace{-40mm}
\caption{\small Baseline {\sf HATRIC} versus approaches with eager
  update of directory on cache and translation structure evictions
  ({\sf EGR-dir-update}), fine-grained tracking of translations ({\sf
    FG-tracking}), and an infinite directory with no
  back-invalidations ({\sf No-back-inv}). {\sf All} comines these
  approach. We show average runtime and energy, normalized to the
  metrics for the best paging policy without {\sf HATRIC}.}
\label{sensitivity}
\vspace{-4mm}
\end{minipage}
}
\end{figure}

\vspace{2mm}\noindent{\bf Coherence directory design decisions:}
Sec. \ref{hardware-design} detailed the nuances modifying traditional
coherence directories to support translation coherence. Figure
\ref{sensitivity} captures the performance and energy (normalized to
those of the best paging policy or {\sf sw} in previous graphs) of
these approaches. We consider the following options, beyond baseline
{\sf HATRIC}.

\vspace{2mm}\noindent{\sf EGR-dir-update:} This is a design that
eagerly updates coherence directories whenever a translation entry is
evicted from a CPU's L1 cache or translation structures. While this
does reduce spurious coherence messages, it requires expensive lookups
in translation structures to ensure that entries with the same co-tag
have been evicted. Figure \ref{sensitivity} shows that the performance
gains from reduced coherence traffic is almost negligible, while
energy does increase, relative to {\sf HATRIC}.

\vspace{2mm}\noindent{\sf FG-tracking:} We study a hypothetical design
with greater specificity in translation tracking. That is, coherence
directories are modified to track whether translations are cached in
the TLBs, MMU caches, nTLBs, or L1 caches. Unlike {\sf HATRIC}, if a
translation is cached only in the MMU cache but not the TLB, the
latter is not sent invalidation requests. Figure \ref{sensitivity}
shows that while one might expect this specificity to result in
reduced coherence traffic, system energy is actually slightly higher
than {\sf HATRIC}. This is because more specificity requires more
complex and area/energy intensive coherence directories. Further, since
the runtime benefits are small, we believe {\sf HATRIC remains the
  smarter choice}.

\vspace{2mm}\noindent{\sf No-back-inv:} We study an unrealistically
ideal design with infinitely-sized coherence directories which never
need to relay back-invalidations to private caches or translation
structures. We find that this does reduce energy and runtime, but not
significantly from {\sf HATRIC's} dual-grain coherence directory based
on \cite{zebchuk:multi}.

\vspace{2mm}\noindent{\sf All:} Figure \ref{sensitivity} compares {\sf
  HATRIC} to an approach which marries all the optimizations
discussed. {\sf HATRIC} almost exactly meets the same performance and
is actually more energy-efficient, largely because the eager updates
of coherence directories add significant translation structure lookup
energy.

\vspace{2mm}\noindent{\bf Comparison with UNITD:} We now compare {\sf
  HATRIC} to prior work on {\sf UNITD} \cite{romanescu:unified}. To do
this, we first upgrade the baseline {\sf UNITD} design in several
ways. First, and most importantly, we extend add support for
virtualization by storing the system physical address of nested page
tables entries are stored in the reverse-lookup CAM originally
proposed \cite{romanescu:unified}. Second, we extend {\sf UNITD} to
work seamlessly with coherence directories. We call this upgraded
design {\sf UNITD++}.

\begin{figure}[t]
\centering
{
\begin{minipage}[t]{0.48\textwidth}
\centering
\vspace{-4mm}
\epsfig{file=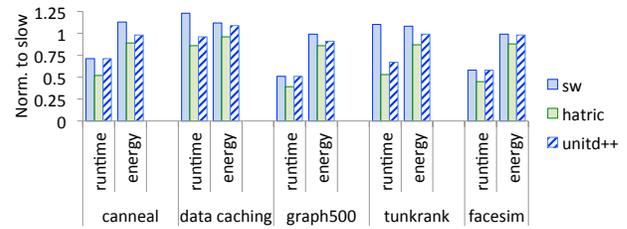, scale=0.34, angle=90}
\vspace{-38mm}
\caption{\small Comparison of {\sf HATRIC's} performance and energy
  versus {\sf UNITD++}. All results are normalized to results for a
  system without die-stacked memory and compared to {\sf sw}.}
\label{unitd}
\vspace{-4mm}
\end{minipage}
}
\end{figure}

Figure \ref{unitd} compares {\sf HATRIC} and {\sf UNITD++} results,
normalized to results from the case without die-stacked DRAM. As
expected, both approaches outperform a system with only traditional
software-based translation coherence ({\sf sw}). However, {\sf HATRIC}
typically provides an additional 5-10\% performance boost versus {\sf
  UNITD++} by also extending the benefits of hardware translation
coherence to MMU caches and nTLBs. Further, {\sf HATRIC} is more
energy efficient than {\sf UNI\-TD++} as it boosts performance (saving
static energy) but also does not need reverse-lookup CAMs.

\vspace{2mm}\noindent{\bf Xen results:} In order to assess {\sf
  HATRIC's} generality across hypervisors, we have begun studying it's
effectiveness on Xen. Because our memory traces require months to
collect, we have thus far evaluated {\sf canneal} and {\sf data
  caching}, assuming 16 vCPUs. Our initial results show that Xen's
performance is improved by 21\% and 33\% for {\sf canneal} and {\sf
  data caching} respectively, over the best paging policy employing
software translation.

\section{Conclusion}
We present a case for folding translation coherence atop existing
hardware cache coherence protocols. We achieve this with simple
modifications to translation structures (TLBs, MMU caches, and nTLBs)
and with state-of-the-art coherence protocols. Our solutions are
general (they support nested and guest page table modifications) and
readily-implementable. We believe, therefore, that {\sf HATRIC} will
become essential for upcoming systems, especially as they rely on page
migration to exploit heterogeneous memory systems.

\bibliographystyle{ieeetr}
\bibliography{references}

\end{document}